\documentclass[11pt]{article}
\usepackage{hyperref}
\usepackage{cite}
\usepackage{graphicx}
\usepackage{subfig}
\usepackage{caption}
\usepackage[top=3cm, bottom=3cm, left=1.5cm, right=1.5cm]{geometry}
\usepackage{amsmath,amsfonts,amssymb,mathtools}
\numberwithin{equation}{section} 
\usepackage{float}
\usepackage{breqn}
\usepackage{multicol}
\usepackage{multirow}
\usepackage{perpage}
\usepackage{booktabs}
\usepackage{xcolor,colortbl}
\definecolor{Gray}{gray}{0.85}
\usepackage{xcolor}
\usepackage{array,makecell, cellspace}
\date{}

\begin{document}
	
	\setcounter{page}{1}
	\pagestyle{plain}

	\begin{center}
		\Large{\bf  Reheating after the Higgs Inflation}\\
		\small \vspace{1cm} {\bf Manda Malekpour\footnote{m.malekpour@stu.umz.ac.ir} \quad and \quad {\bf Kourosh
				Nozari\footnote{knozari@umz.ac.ir(Corresponding Author)}}}
		\\
		\vspace{0.25cm}
		Department of Theoretical Physics, Faculty of Science,\\
		University of Mazandaran,\\
		P. O. Box 47416-95447, Babolsar, Iran
	\end{center}
	
	\begin{abstract}
		
		Since the discovery of the Higgs particle in the LHC, numerous inflationary models utilize it as the inflaton. In this study, after establishing that Higgs inflation can be a viable model in the context of unimodular gravity, we address a scenario related to reheating, which plays a crucial role in connecting the inflationary era to a hot Big Bang universe. Also, we calculate essential parameters such as the temperature and the e-folds number according to the equation of the state parameter ($\omega_{re}$) in this regime of reheating. The physics behind the reheating process are extremely imprecise at present, and we have only a very limited range of acceptable values for $\omega_{re}$. We analyze the compatibility with Planck2018, BICEP/Keck2021, DESI2024, and ACT array bounds and find that within this model, $\omega_{re}$ is restricted to the range $ -1 \leq \omega_{re} \leq \frac{1}{6} $.\\

	\end{abstract}

	\newpage	
	\section{Introduction}
	
	After decades of cosmological explorations, cosmologists found that the flat $\varLambda CDM$ model is successful in explanation of the universe evolution in many respects, from the Big Bang to the late time cosmic speed up \cite{A.D. Linde,Rio02,baumann2018tasi,Mishra:2024axb}.
	After Alan Guth introduced the concept of \textit{\textquotedblleft cosmological inflation\textquotedblright} in 1981 \cite{guth1981inflationary}(usually referred to as the old inflation model), a new scenario introduced a canonical scalar field as the inflaton (usually referred to as the new inflation model) \cite{linde1983chaotic} that slowly rolls towards the minimum of its effective potential and successfully derives the inflation stage.
	A primary observation in cosmology shows that a short period of rapid cosmic expansion (cosmic inflation), provides a framework to explain problems of the hot Big Bang cosmology and the formation of the large scale structure in the universe \cite{Senatore:2016aui}. In fact, quantum fluctuations in an inflation stage created the seeds for the structure formation. The fingerprints of such early time perturbations now are detectable in the Cosmic Microwave Background (CMB) radiation temperature maps.
	
	Although the nature of the scalar field remains unclear, the Higgs boson as the only fundamental scalar field is the simplest candidate for the inflaton observed to date \cite{Sadjadi:2013na}. Specifically, both the minimally and non-minimally coupled Higgs inflation are studied in the literature \cite{Bezrukov:2007ep, He:2023vlj}. The non-minimal coupling (NMC) with gravity triggers inflation and addresses several issues such as unitary, suppressed tensor-to-scalar ratio and fine-tuning \cite{Faraoni:2000wk}. Thus, considering that the NMC parameter $ \xi $ is regarded as inevitable in many inflationary scenarios, it has been the focus of numerous studies in recent decades \cite{Makino:1991sg,M.De,faraoni1996nonminimal,faraoni1998conformal,faraoni2000inflation,park2008inflation,hertzberg2010inflation,qiu2012reconstruction,
shaposhnikov2013cosmology,Nozari2012,Nozari2013,chiba2015consistency,Myrzakul:2015gya,bostan2019quartic}.
	
	While the cosmological constant has been introduced for a century ago, its origin and nature is still an open question in standard model of particle physics and also cosmology. A possible alternative for addressing certain issues in the standard cosmology is the framework of \textquotedblleft\textit{Unimodular Gravity}\textquotedblright (UG). In 1919, Einstein wrote the first trace-free field equations \cite{einstein1952principle}. This formulation suggests that the cosmological constant can be considered as an integration constant in the field equations. In recent decades, various models for UG, as a modified theory of gravity, have been constructed \cite{gao2014cosmological,cho2015unimodular,Nojiri:2015sfd,Nojiri:2016ppu,Nojiri:2016plt,barvinsky2017darkness,barvinsky2019inflation,barvinsky2019dynamics}.
	UG is used for additional restriction that the determinant of the metric is fixed to be unity or as a function of the spatial coordinates- most commonly as $\sqrt{-g}=1$.
	So, obviously this restriction imposes constraints on the variations derived from the Einstein-Hilbert action and leads to field equations that differ from those in GR.
	Therefore, this situation provides a different viewpoint about the cosmological constant; nonetheless, the physical results are similar to GR's.
	As a result, within this theoretical framework, the cosmological constant operates as a Lagrange multiplier to drive cosmological inflation \cite{weinberg1989cosmological,unruh1989unimodular,Henneaux:1989zc,liddle1993end,sahni2002cosmological}.
	
		Of course, there was no connection of Einstein's research to the cosmological constant, but it was about the structure of point particles in GR.
		Then, in 1970, two physicists named Anderson and Finkelstein \cite{anderson1971cosmological} discussed and argued that it is natural to decompose the metric into two separate parts: one being the determinant of the metric $g$, and the remaining metric constrained to have unit determinant.
		To provide a mechanism for accelerating the universe, the simplest candidate can be the vacuum energy density from particle physics, that is, the cosmological constant. Interestingly, the present expansion of the universe can already be explained by considering only a single component, such as the cosmological constant or non-relativistic matter \cite{jain2012testing,jain2012cosmological}.
		Although the cosmological constant alone cannot trigger cosmic inflation, it can be regarded as a candidate for the dark energy, which has caused the recent positively accelerated expansion of the universe.
		According to quantum field theory, the cosmological constant arising from the vacuum expectation values of quantum fields, is predicted to exceed the observed value by about $60$ to $120$ orders of magnitude \cite{nojiri2017modified}. One of the reasons for considering UG is to address the cosmological constant problem.
		While the cosmological constant was manually added to the standard Einstein field equations, in UG it emerges by imposing a constraint on the determinant of the metric.
		This theory is constructed in such a way that the determinant of the spacetime metric is not dynamical, but restricted.
		This is a huge success for UG, because a cosmological constant arises naturally from the trace-free Einstein equations as an integration constant. Such a constant may compensate the large value of the vacuum energy density. Therefore, this approach potentially resolves the initial fine-tuning problem of the cosmological constant.
	
	A phenomenologically rich and complicated stage of the post-inflationary era is the \textit{reheating} era. The main point here is that during the slow roll evolution, the scalar field (for instance, the Higgs field) slowly rolls down toward its minimum of the potential and begins to oscillate around the local minimum. Once inflation has ended, the inflaton field decays into radiation and reheats the universe. So, the energy density of the inflaton field finally converts into other particles/radiation at the end of inflation \cite{Cai:2015soa, Cook:2015vqa, Munoz:2014eqa, Lozanov:2019jxc}. This means the reheating process describes translating from the inflationary epoch to the next evolution stage of the universe, leading the radiation-dominated universe \cite{Pareek:2021lxz, Amin:2014eta}. In this approach, there is the idea that the dark matter (for instance via the formation of the primordial black holes) and the primordial gravitational waves could generate from the reheating process. Hence,  there are several important parameters in the reheating era such as the reheating e-folds number $ N_{re} $, the reheating temperature $T_{re}$ (at the beginning of the radiation-dominated era), and the equation of state $\omega_{re}$ that are crucial to be studied in order to understand the nature of reheating stage. The existing literature show that using the CMB data, the upper bound on the reheating temperature $T_{re}$ can reach approximately $ \sim  10^{16}\,\mathrm{GeV}$, while the lower bound is set by Big Bang Nucleosynthesis (BBN) at $T_{re} \gtrsim 10\,\mathrm{GeV} $ \cite{Cook:2015vqa}.
	
	In the present study, following our previous work \cite{Nozari:2024hip}, we study reheating process in a Higgs inflation model in the framework of UG. We begin with a general overview of the inflation paradigm in this setup. Then our main goal is to investigate the reheating phase of the universe in this framework and obtain corresponding observable parameters for Higgs inflation. Furthermore, we analyze the predictions of our model and constrain them with observational data from the Planck2018, BICEP/Keck(BK)2021, DESI2024, and ACT datasets at $68\%$ and $95\%$ levels of confidence \cite{akrami2020planck, Planck:2018vyg, ade2021improved, Paoletti:2022anb, daCosta:2024grm, ACT:2025tim, ACT:2025fju}. In this manner we find a distinctive range of the equation of state parameter of the cosmic fluid at the reheating stage.
	
	The structure of this paper is organized as follows: In section \eqref{sec-2}, we provide a brief overview of the Higgs inflationary dynamics within the framework of UG. Then, we calculate the important key inflationary parameters in this setup. Section \eqref{sec-3} is devoted to the computation of $ T_{re} $ and $ N_{re} $ for the Higgs inflation in UG for various choices of the equation of state parameter $ \omega_{re} $. In each step we compare our results with the most recent observational data from  Planck2018, BICEP/Keck2021, DESI2024, and ACT datasets. In section \eqref{sec-4}, we summarize our main results. Throughout this work, we adopt the spacetime metric signature as $(-, +, +, +)$ for the sign convention and for units we set $ c = \hbar=1$, $\kappa^{2}\equiv 8 \pi G = M_{P}^{-2} $.
	
	
	\section{The Higgs inflation in unimodular gravity}\label{sec-2}
	
\begin{figure}[H]
		\centering
		\includegraphics[height= 10cm, width=12cm]{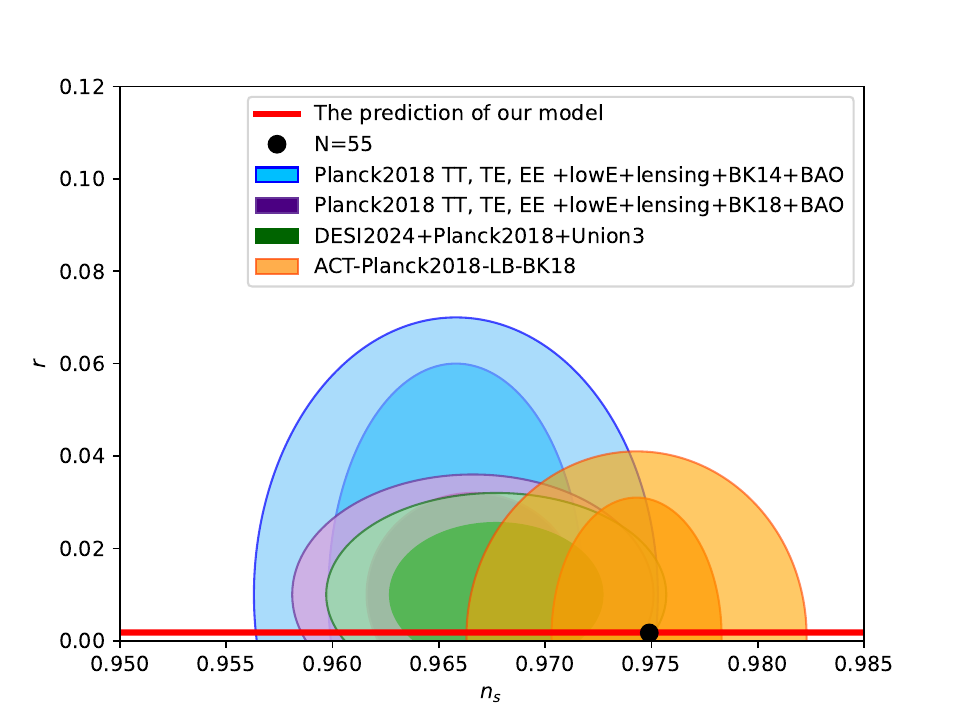}
		\caption{Constraints on the $r$ - $n_{s}$ parameter space at the $ 68\% $ and $ 95\% $ CL, based on the Planck 2018, BICEP/Keck(BK)2021, DESI2024, and ACT observational data. The red line corresponds to the relationship between $r $ - $ n_{s}$, while the yellow bullet marks the value corresponding to $ N=55 $ within the Higgs inflation model.}\label{fig.11}
	\end{figure}

	During the Higgs inflation, the inflaton field has a significant NMC to gravity. In this section, we provide a comprehensive summary of the Higgs inflationary model, touching on its main aspects and fundamental concepts (for more details, see Ref. \cite{Nozari:2024hip}).
	One of the models that proposes to solve the cosmological constant problem is UG. As we discussed in the previous section, the basic principle of UG is centered on limiting the metric determinant value, $\sqrt{-g}=1$.
	For a homogeneous and isotropic space-time, the Friedmann-Robertson-Walker (FRW) metric is defined as follows
	\begin{equation}\label{eq1}
		ds^2 = -dt^{2} + a^{2}\left(t\right) \sum_{i=1}^{3} \left( dx^{i}\right)^{2} \,,
	\end{equation}
	where $a\left( t\right)$ denotes the scale factor. Given that the above equation violates the unimodular constraint, we address this issue by considering the following approach
	\begin{equation}\label{eq2}
		d\tau = a^{3}\left(t\right) dt\,.
	\end{equation}
	In light of this, the FRW metric may be expressed in the following form
	\begin{equation}\label{eq3}
		ds^{2} = -a^{-6} \left(\tau\right)  d\tau^{2} + a^{2}\left(\tau\right)  \sum_{i=1}^{3} \left(dx^{i}\right)^{2} \,.
	\end{equation}
By using the metric \eqref{eq3}, the non-zero components of the Ricci tensor and scalar as follows   
	\begin{equation}\label{eq14}
		\begin{split}
			R_{\tau\tau}= -3\dot{\mathcal{H}} - 12 \mathcal{H}^2\,,
			\quad R_{ij}= a^8\left(\dot{\mathcal{H}}+ 6\mathcal{H}^2\right)  \delta_{ij}\,,
			\quad R= a^6 \left(6\dot{\mathcal{H}} +30 \mathcal{H}^2\right) \,,
		\end{split}
	\end{equation}
	where Hubble rate define as $\mathcal{H}=\frac{1}{a}\frac{da}{d\tau}$.
In Ref. \cite{Nozari:2024hip}, we examined and discussed the primary cosmological parameters related to the Higgs field formulated within the framework of UG and developed this model specifically in the UG context. The action is constructed with two sections: the NMC in UG and the SM Higgs ($ S_{J}= S_{\phi} + S_{SM}$).
	In this model, the action is considered to be
	\begin{equation}\label{eq4}
		S_{J}=\int d^{4}x \left\lbrace \sqrt{-g} \left(f\left( \phi \right) R-\frac{1}{2} g^{\mu \nu} \partial_{\mu}\phi \partial_{\nu}\phi - V\left( \phi\right)\right) - \lambda \left(\phi\right)\left( \sqrt{-g}-1\right) \right\rbrace\,,
	\end{equation}
	where $ \lambda $ is the Lagrange multiplier function and $ V\left( \phi\right)= \frac{\gamma}{4} \left(\phi^{2} -\upsilon^{2}\right)^{2} $ is the potential of the Higgs field with $\upsilon$ as the vacuum expectation value. In this formulation, $ f\left(\phi \right) $ represents a scalar field that interacts non-minimally with gravity through a coupling to the Ricci scalar $ R $ in the following form
	\begin{equation}\label{eq5}
		f\left(\phi \right)=\frac{M_{P}^{2}}{2} \left(1+\frac{\xi \phi^2}{M_{P}^{2}}\right)\,,
	\end{equation}
	where $\xi $ quantifies the strength of the non-minimal coupling, and $M_{P} $ denotes the reduced Planck mass. Next, we find the field equations as follows
	\begin{equation}\label{eq6}
		-\frac{1}{2} g_{\mu\nu} f\left(\phi\right) R + f\left(\phi\right)  R_{\mu\nu} +
		\left( \Box g_{\mu\nu}- \nabla_{\mu} \nabla_{\nu} \right) f\left( \phi\right) + \frac{\left(\lambda \left(\phi\right)+V\left( \phi\right) \right)}{2}  g_{\mu\nu} + \frac{1}{4}g_{\mu\nu} \nabla_{\alpha}\phi \nabla^{\alpha}\phi - \frac{1}{2}\nabla_{\mu}\phi \nabla_{\nu}\phi=0 \,.
	\end{equation}
	We can find the field equations as follows
	\begin{equation}\label{eq7}
		3 \mathcal{H}^{2} f\left(\phi\right)+ 3 \mathcal{H} \dot{f}\left(\phi\right) - \left( \frac{\lambda \left(\phi\right)+V\left( \phi\right)}{2} \right) a^{-6}- \frac{1}{2} \dot{\phi^{2}}=0\,,
	\end{equation}
	\begin{equation}\label{eq8}
		\left(-2 \dot{\mathcal{H}} -9 \mathcal{H}^{2}\right) f\left(\phi\right)-5 \mathcal{H} \dot{f}\left(\phi\right)- \ddot{f}\left(\phi\right)+\left(\frac{\lambda \left(\phi\right)+V\left( \phi\right)}{2}\right) a^{-6}-\frac{1}{2} \dot{\phi^{2}}=0\,,
	\end{equation}
	where dot over a quantity marks it's time derivative as \textquotedblleft $ \cdot $\textquotedblright = $\frac{d}{d\tau}$. Next, the energy-momentum tensor corresponding to Eq. \eqref{eq6} is given by
		\begin{equation}\label{eq29}
			\left( R_{\mu \nu}-\frac{1}{2} g_{\mu \nu} R\right) =- \frac{1}{f\left(\phi\right)} \left( \left(g_{\mu \nu} \Box - \nabla_{\mu} \nabla_{\nu}\right) f\left(\phi\right) +\left(\frac{\lambda\left( \phi\right)+V\left( \phi\right)}{2}\right) g_{\mu \nu}- \frac{1}{2}g_{\mu\nu} \nabla_{\alpha}\phi \nabla^{\alpha}\phi\right)  \,.
	\end{equation}
	Then by employing the equation $G_{\mu \nu}\equiv R_{\mu \nu}-\dfrac{1}{2} g_{\mu \nu} R = \frac{T^{(eff)}_{\mu \nu}}{M_{p}^{2}}$, we find the energy-momentum tensor in this setup as follows
	\begin{equation}\label{eq9}
		T_{\mu \nu}^{(eff)} \equiv - \frac{M_{P}^{2}}{f(\phi)} \left(\left(g_{\mu \nu} \Box - \nabla_{\mu} \nabla_{\nu}\right) f(\phi)+\left(\frac{\lambda\left( \phi\right)+V\left( \phi\right)}{2}\right) g_{\mu \nu}+ \frac{1}{2}g_{\mu\nu} \nabla_{\alpha}\phi \nabla^{\alpha}\phi\right)\,.
	\end{equation}
	Through the application of the continuity equation
	\begin{equation}\label{eq10}
		\dot{\rho}_{_{eff}} + 3 \mathcal{H} \left( \rho_{_{eff}} + P_{_{eff}} \right) =0 \,,
	\end{equation}
	and using the equations above, we obtain expressions for the effective energy density $\rho_{_{eff}}\left(\phi\right)$ and pressure $P_{_{eff}}(\phi)$ for this model, which can be defined as follows
	\begin{equation}\label{eq11}
		\begin{split}
			& \rho_{_{eff}}\left( \phi\right)  = - \frac{M_{P}^{2}}{f\left( \phi\right)} \left[3 \mathcal{H} \dot{f}\left( \phi\right) a^{6} - \left( \frac{\lambda\left( \phi\right)+V\left( \phi\right)}{2}\right) -\frac{1}{2}\dot{\phi}^{2}a^{6}\right]\,,\\
			& P_{_{eff}}\left( \phi\right)  = -\frac{M_{P}^2}{f(\phi)} \left[ \left( -\ddot{f}(\phi)- 5 \mathcal{H} \dot{f}(\phi) \right) a^{6}+\left( \frac{\lambda\left( \phi\right)+V\left( \phi\right)}{2}\right) -\frac{1}{2}\dot{\phi}^{2} a^{6}\right]\,.
		\end{split}
	\end{equation}
	By using Eqs. \eqref{eq10} and \eqref{eq11}, we find
	\begin{equation}\label{eq12}
		\begin{split}
			&\left( \frac{f(\phi)-\frac{3}{2}\xi^{2} \phi^{2}}{\xi \phi}\right) \ddot{\phi}+3 \mathcal{H} \dot{\phi} \left[\frac{3 \xi^{2} \phi^{2}+ 2f(\phi)}{\xi \phi} \right]  +\frac{3}{4} \dot{\phi}^{2}+ \frac{3}{2} \xi \dot{\phi}^{2}
			-\left( V\left(\phi\right) + \lambda\left(\phi\right) \right) a^{-6}\\
			& +\frac{\left(\lambda^{\prime}\left( \phi\right) +V^{\prime}\left(\phi \right) \right) f(\phi) a^{-6}}{2\xi \phi}=0\,.
		\end{split}
	\end{equation}
where $ \lambda^{\prime}\left( \phi\right) $ is defin as $ \lambda^{\prime}\left( \phi\right) = \frac{d\lambda}{d\phi } $. From this point onward, we will omit the explicit dependence of $\lambda$ and $V$ on $\phi$ for the sake of brevity.
	To proceed further, during the slow-roll stage, we have $ \frac{d^2 \phi}{d\tau^{2}} \ll -2 \mathcal{H} \frac{d \phi}{d\tau},\,\,  \frac{d \phi}{d\tau}\ll \mathcal{H} \phi,\,\,
	\frac{d\mathcal{H} }{d\tau}\ll -2 \mathcal{H}^{2},\,\, \left( \frac{d\phi}{d\tau}\right)^{2}  \ll V$, and the equations of motion become
	\begin{equation}\label{eq13}
		\mathcal{H}^{2} \simeq\frac{2 \left( \lambda+V\right)  a^{-6}}{3 M_{P}^{2} \left( 1+\dfrac{\xi \phi^{2}}{M_{P}^{2}}\right)} \,,	
	\end{equation}
	\begin{equation}\label{eq14}
		3\mathcal{H} \dot{\phi} \simeq \frac{1}{\left( \xi \phi^{2}\left(1+3\xi \right)+M_{P}^{2}\right)}\left( \left( \left( \lambda+V\right)a^{-6}\right) -\frac{\left( \lambda^{\prime}+V^{\prime}\right) \left(1+\frac{\xi\phi^{2}}{M_{P}^{2}}\right) M_{P}^{2}a^{-6}}{4\xi\phi}\right)\,.
	\end{equation}
	Furthermore, using the definitions of the slow-roll parameters $ \epsilon_{1} $, $ \epsilon_{2} $, $ \epsilon_{3} $ and $ \epsilon_{4} $ in terms of $ \mathcal{H} $, we obtain the following
	\begin{equation}\label{eq15}
		\epsilon_{1}= -3 - \frac{\dot{\mathcal{H}}}{\mathcal{H}^{2}}\,, \quad
		\epsilon_{2}= 3+\frac{\ddot{\phi}}{\mathcal{H} \dot{\phi}}\,, \quad
		\epsilon_{3}= \frac{\dot{f}\left( \phi\right)}{2 \mathcal{H} f\left( \phi\right) }\,, \quad
		\epsilon_{4}= \frac{\dot{E}}{2 \mathcal{H} E}\,,
	\end{equation}
	where by definition
	\begin{equation}\label{eq16}
		E\equiv f\left(\phi\right)+ \frac{3 \dot{f}^{2}\left(\phi\right)}{2 \kappa^{2} \dot{\phi}^{2}} = 1 + \left(1 + 6\xi\right) \frac{\xi\phi^{2}}{M_{P}^{2}}\,,
	\end{equation}
and
	\begin{equation}
			\dot{E}=  \frac{2 \left( 1 + 6 \xi \right) }{M_{P}^{2}} \dot{f}\left( \phi\right) \,.
	\end{equation}	
	This leads to the determination of the observable quantities $r $ and $ n_{s} $, expressed as
	\begin{equation}\label{eq17}
		n_{s} \simeq 1- 4 \epsilon_{1} - 2 \epsilon_{2} +2\epsilon_{3} -2 \epsilon_{4}\,,
	\end{equation}
	\begin{equation}\label{eq18}
		r= 8 \kappa^{2} \frac{Q_{s}}{f\left( \phi\right)}\,,
	\end{equation}
	where $Q_{s}$ is defined as follows
	\begin{equation}\label{eq19}
		Q_{s}\equiv \frac{E \dot{\phi^{2}} }{f\left( \phi\right)  H^{2} (1 +\epsilon_{3} )^{2}}\,.
	\end{equation}
	Additionally, the number of e-folds, $ N_{k} $, from the departure of a mode with wavenumber $ k $ from the horizon up to the end of inflation is given by the relation
	\begin{equation}\label{eq111}
		N_{k}= \int_{t_{k}}^{t_{e}} H \, dt = \int_{\phi_{k}}^{\phi_{e}} \frac{H}{\dot{\phi}} \, d\phi\,,
	\end{equation}
	where, the subscripts \textquotedblleft $ k $\textquotedblright and \textquotedblleft e \textquotedblright refer to the quantities at the horizon exit of the mode and at the end of inflation, respectively. In the \eqref{appendix}appendix, we provide additional details and calculations. In order to determine the observationally viable parameters space of the model, for the coupling constant $\xi $ we impose the constraint on $n_{s}$ such that the calculated value is $\xi=6 \times 10^{2} $, corresponding to $N= 55$ e-folds. This value corresponds to the spectral index $n_s= 0.9752$ and tensor-to-scalar ratio $r=0.0017$ which are in excellent agreement with the constraints provided by Planck2018, BICEP/Keck2021 and DESI2024 data sets. These points will be discussed in the next section where we derive the relationship between the inflationary parameters and the reheating parameters, $ T_{re} $ and $ N_{re}$. Figure \eqref{fig.11} illustrates the $r$ - $n_{s}$ trajectory predicted by the Higgs inflation model, overlaid with the Planck2018 TT, TE, EE +lowE+lensing+BK\textbf{14}(\textbf{18})+BAO, in the context of $ \varLambda CDM+r+\frac{dn_{s}}{dn_{k}}$ constraints at both $68\%$ and $95\%$ confidence levels (CL). We note that according to recent observational data, the scalar spectral index is $n_{s}=0.9658\pm 0.0038$, while the tensor-to-scalar ratio is constrained to $r<0.072$ with BICEP/Keck\textbf{14} and $r<0.036$ BICEP/Keck\textbf{18}. Also, a data combination of DESI2024+CMB(Planck2018)+Union3 gives an upper bound on the tensor-to-scalar ratio $r<0.032$, and the spectral index $n_{s}=0.9673\pm 0.0036$. In comparison with the latest ACT+Planck2018+LB+BK18 data, in which the tensor-to-scalar ratio is constrained to $r<0.038$ at $95\%$ CL and the spectral index is $n_{s}=0.9743\pm 0.0034$, our model prediction remains consistent within the confidence level.\\
	
	\section{Reheating after the Higgs inflation in unimodular gravity}\label{sec-3}

\begin{figure}[H]
		\centering
		\includegraphics[height= 5.5cm, width=11cm, width=0.45\textwidth]{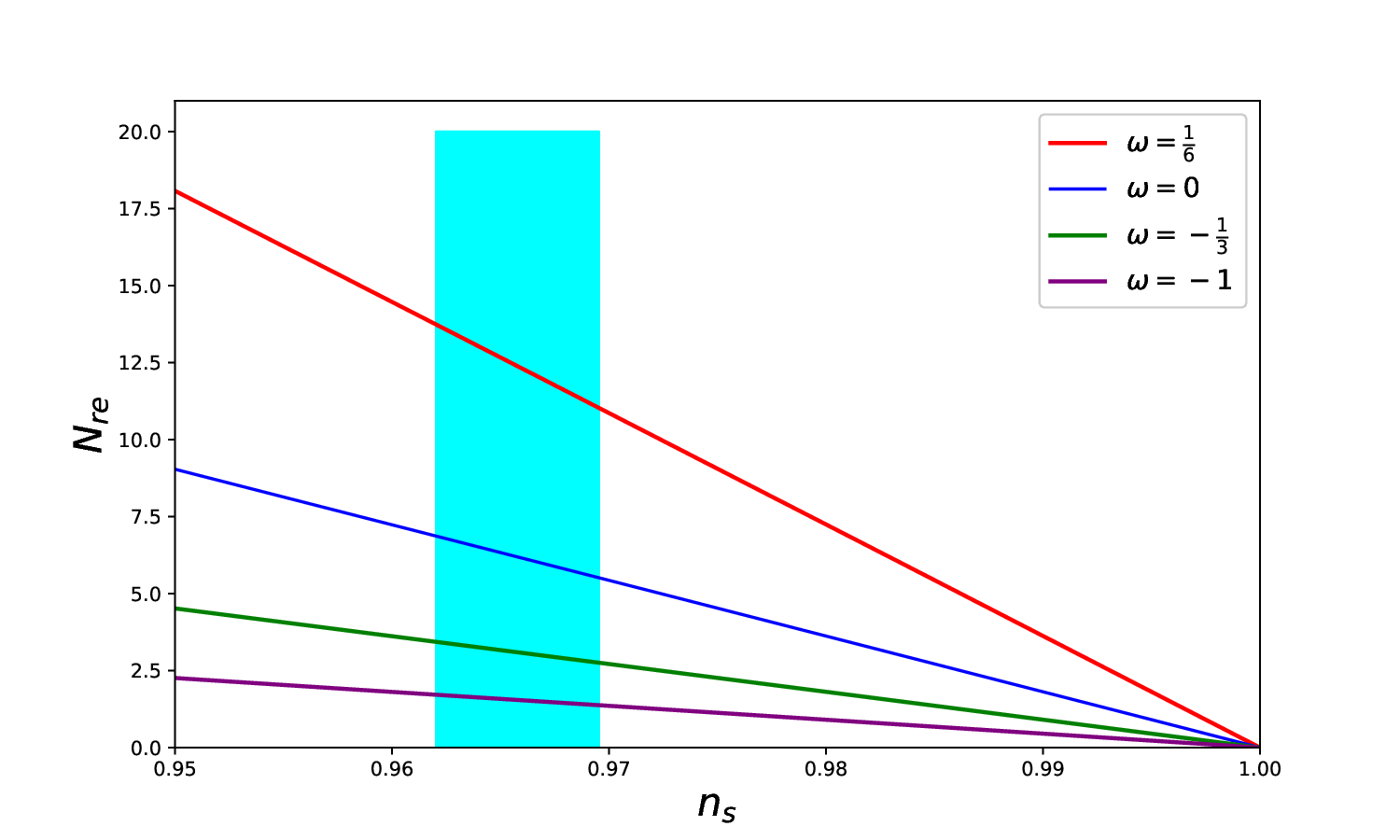}
		\hfill
		\includegraphics[width=0.45\textwidth]{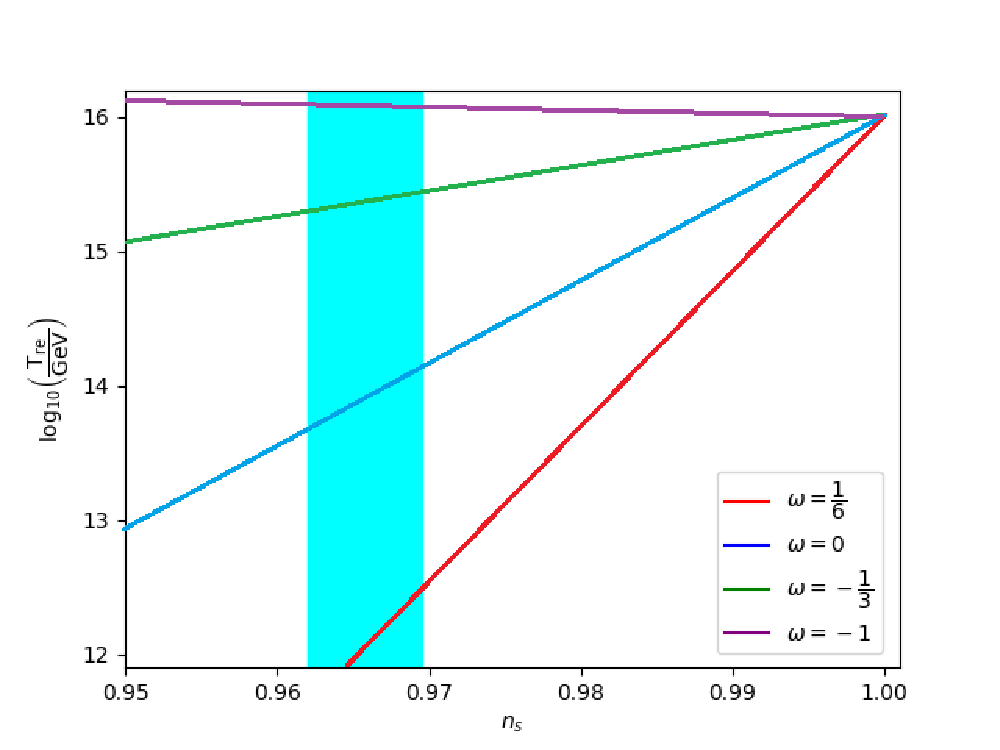}
		\caption{Left Panel: Variation of \( N_{re} \) with respect to \( n_s \) for different choices of \( \omega_{re} \), within the framework of UG Higgs inflation. The red, blue, green, and purple lines correspond to \( \omega_{re} = \frac{1}{6}, 0, -\frac{1}{3}, \) and \(-1\), respectively. The vertical cyan strip represents the bounds on $n_{s}$. Right Panel: Final reheating temperature, $ \log_{10}\!\left(\frac{T_{\mathrm{re}}}{\mathrm{GeV}}\right)$, plotted as a function of the scalar spectral index \( n_s \) for four different values of the equation of state parameter \( \omega_{re} \).}
		\label{fig.2}
	\end{figure}

\begin{figure}[H]
		\centering
		\includegraphics[height= 8cm, width=10cm]{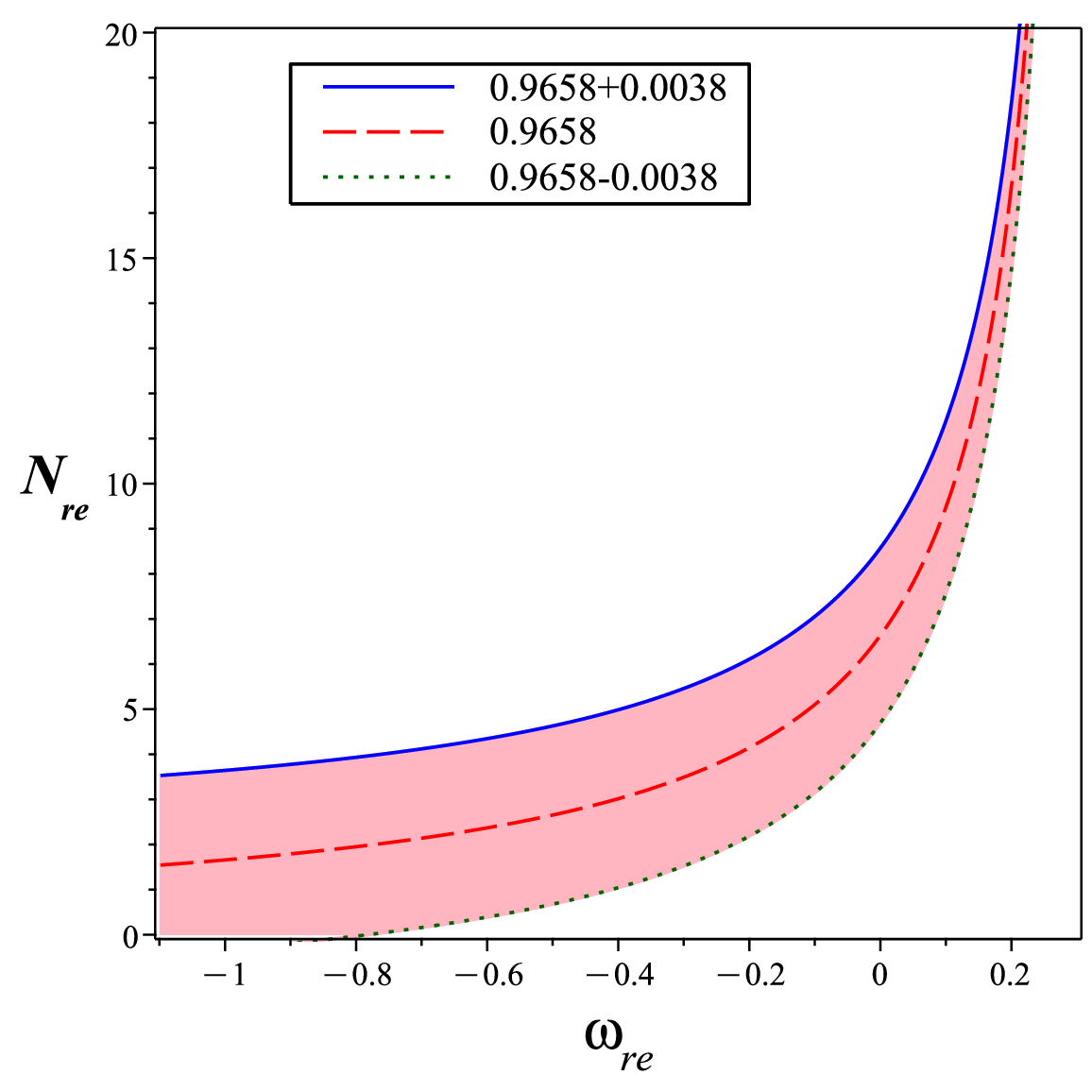}
		\caption{Variation of the number of e-folds during reheating \( N_{re} \) with respect to the equation of state parameter \( \omega_{re} \).
		}\label{fig.3}
	\end{figure}

	As we mentioned in the Introduction, the reheating phase is an important stage in transitioning from the inflationary epoch to a hot Big Bang universe. After the end of the inflationary phase, the energy density of the inflaton field was transferred to the elementary particles content of the SM, and then the thermal history of the universe began \cite{Pareek:2021lxz}. We carry out a comprehensive analysis of reheating in the Higgs inflation and calculate reheating parameters including ($T_{re}$, $N_{re}$, and $\omega_{re}$). For this purpose, the number of e-folds is defined as
	\begin{equation}
		N_{k} = \ln\left( \frac{a_{end}}{a_{k}}\right)\,,
	\end{equation}
	where $ a_{k} $ represents the scale factor at the time of horizon exit, and $ a_{end} $ denotes the scale factor at the end of inflation.
	In this section, following the approach of Refs. \cite{Cook:2015vqa, Munoz:2014eqa, Lozanov:2019jxc}, we assume a constant equation of state during the reheating phase. By applying the relation between the scale factor and the energy density, $\rho \propto a^{-3(1+\omega)}$, the evolution of the energy density in the reheating period can then be stated as follows
	\begin{equation}\label{eq21}
		\frac{\rho_{end}}{\rho_{re}} = \left(\frac{a_{end}}{a_{re}} \right)^{-3\left(1+\omega_{re} \right) }\,,
	\end{equation}
	where the subscript \textquotedblleft end\textquotedblright  signifies the termination of the inflationary phase and reheating start, while \textquotedblleft re\textquotedblright represents the reheating era. Based on Eq. \eqref{eq21}, the expansion e-folds during the reheating phase can be formulated as
	\begin{equation}\label{eq22}
		N_{re} = \ln\left( \frac{a_{re}}{a_{end}}\right) = \frac{1}{3 \left(1+\omega_{re} \right) } \ln\left(\frac{\rho_{end}}{\rho_{re}} \right)\,.
	\end{equation}
	Once reheating ends, the universe smoothly enters the radiation-dominated era, marking the beginning of its thermal history as described by the standard cosmological model. This leads to the following expression for the radiation energy density in terms of the reheating temperature
	\begin{equation}\label{eq23}
		\rho_{re} = \frac{\pi^{2}}{30} g_{re} T^{4}_{re}\,,
	\end{equation}
	where $g_{re}$ is the effective number of relativistic species present at the reheating stage. We set $ g_{re}=106.76$ and adopt the Planck's pivot scale of $ 0.05\, \mathrm{Mpc^{-1}}$ in our forthcoming discussions. Combining Eqs. \eqref{eq22} and \eqref{eq23} allows us to express $  N_{re} $ explicitly in terms of the reheating temperature $ T_{re} $ as follows 
	\begin{equation}\label{eq24}
		N_{re} = \frac{1}{3 \left(1+\omega_{re} \right) }\ln\left(\frac{30 \rho_{_{end}}}{\pi^{2} g_{re} T^{4}_{re}} \right)\,.
	\end{equation}
	By considering the fact that entropy remains conserved after reheating to the present epoch, one can establish a relation between the reheating temperature and the current CMB temperature as follows
	\begin{equation}\label{eq25}
		T_{re} = T_{0} \left( \frac{a_{0}}{a_{re}}\right) \left( \frac{43}{11 g_{re}}\right)^{1/3} = T_{0} \left( \frac{a_{0}}{a_{eq}}\right) e^{N_{_{RD}}} \left( \frac{43}{11 g_{re}}\right)^{1/3}\,,
	\end{equation}
	where the subscript \textquotedblleft 0\textquotedblright  indicates that the corresponding quantities are evaluated at the present time, the radiation-matter equality is marked by \textquotedblleft eq\textquotedblright, and the radiation-dominated
	epoch is specified by \textquotedblleft RD\textquotedblright where $ e^{-N_{_{RD}}} \equiv \frac{a_{re}}{a_{eq}}$. The ratio $\frac{a_{0}}{a_{eq}}$ can now be expressed as
	\begin{equation}\label{eq26}
		\frac{a_{0}}{a_{eq}} = 	\frac{a_{0}}{a_{k}} \frac{a_{k}}{a_{end}} 	\frac{a_{end}}{a_{re}} 	\frac{a_{re}}{a_{eq}}= \frac{a_{0}H_{k}}{k} e^{-N_{k}}  e^{N_{re}} e^{-N_{RD}}\,.
	\end{equation}
	Here, $ H_{k} $ represents the Hubble parameter at the moment of horizon crossing. Moreover, the horizon crossing condition is given by $k = a_{k} H_{k}$.
	By inserting Eq. \eqref{eq26} into Eq. \eqref{eq25}, the following expression is derived
	\begin{equation}\label{eq27}
		T_{re} = \left( \frac{43}{11 g_{re}}\right)^{1/3} \left(\frac{T_{0} a_{0}}{k}\right)  H_{k} e^{-N_{k}} e^{N_{re}}\,.
	\end{equation}
	where $ T_{0} = 2.725\, \mathrm{K} $ as the current CMB temperature.
	Next, by assuming $\omega_{re} \neq \frac{1}{3} $ and by inserting Eq. \eqref{eq27} into Eq. \eqref{eq24}, the number of e-folds during reheating can be formulated as follows 
	\begin{equation}\label{eq28}
		N_{re} = \frac{4}{1- 3 \omega_{re}} \left[- \frac{1}{4} \ln \left( \frac{30}{\pi^{2} g_{re} }\right) - \frac{1}{3} \ln\left( \frac{11 g_{re}}{43}\right) - \ln\left(\frac{k}{T_{0} a_{0}} \right) - \ln\left( \frac{\rho_{end}^{1/4}}{H_{K}}\right) - N_{k}  \right]\,.
	\end{equation}
	Assuming the pivot scale $ k = 0.05\, \mathrm{Mpc^{-1}}$ and taking $ g_{re} = 106.75 $, the expression in Eq. \eqref{eq28} simplifies as follows
	\begin{equation}\label{eq30}
		N_{re} = \frac{4}{1- 3 \omega_{re}} \left[ 61.6 - \ln\left( \frac{\rho_{_{end}}^{1/4}}{H_{K}}\right) - N_{k}\right]\,.
	\end{equation}
	At this stage, employing Eqs. \eqref{eq21} and \eqref{eq22}, we derive an explicit expression for $ T_{re} $
	\begin{equation}\label{eq31}
		T_{re} = \left(\rho_{_{end}} \right)^{1/4} \left[ \frac{30}{\pi^{2} g_{re}}\right]^{1/4} \mathrm{exp}\left[ -\frac{3}{4} N_{re} \left( 1+\omega_{re} \right) \right]\,.
	\end{equation}
	Next, the Hubble parameter $ H_{k} $ can be computed using the definition of the tensor-to-scalar ratio
	\begin{equation}\label{eq32}
		r = \frac{P_{h}}{P_{\zeta}} = \frac{2 H_{k}^{2}}{\pi^{2} M_{P}^{2} A_{s}}\,,
	\end{equation}
	where we adopt $ P_{h}= \left( 2 H_{k}^{2}\right) /\left( \pi^{2} M_{P}^{2}\right)$ and $ P_{\zeta} = A_{s} $. Therefore, by using Eq. \eqref{eq18} and $ H_{k} = a^{3}  \mathcal{H}_{k} $, we find the Hubble parameter during inflation as follows
	\begin{equation}
		\mathcal{H}^{2}_{k} = \frac{1}{a^{6}} \left[ \frac{\pi^{2} M_{P}^{2} A_{s}}{54  \xi^{6} \phi_{k}^{2} }\right]^{1/3}\,.
	\end{equation}
	By applying the field equation \eqref{eq13}, the potential and the Lagrange multiplier at the end of inflation are given by
	\begin{equation}
		V_{end} \simeq \lambda_{end}= \left[\frac{3}{2} \left(\frac{\pi^{2} M_{P}^{2} A_{s}}{54 \phi_{k}^{2}}\right)^{1/3} \left( \frac{\phi_{k}^{2}}{\xi}\right) - \frac{\phi_{k}^{4}}{4}\right] \left( \frac{\phi_{e}^{4}}{\phi_{k}^{4}}\right)\,.
	\end{equation}
	In this model, the potential energy function is given by $ V \simeq \gamma \phi^{4}/4 $, and the Lagrange multiplier $ \lambda\simeq \phi^{4}/4 $. For the purpose of simplification, we adopt the following assumptions: $ M_{P}=1 $, $ \left(1+\gamma \right)\approx 1  $ (that is, $ \gamma\ll 1 $), $ \xi \gg 1 $, and $ A_{s}= 2.196 \times 10^{-9} $. In conclusion, applying Eq.~\eqref{eq11} at the end of inflation under the slow-roll approximation—while neglecting the third equation—yields the following expression for the energy density
	\begin{equation}
		\rho_{_{end}} = \frac{M_{P}^{2}}{6 \phi_{e} \xi^{3}}+\frac{2 M_{P}^{2}}{\xi} \left[\frac{3}{2} \left(\frac{\pi^{2} M_{P}^{2} A_{s}}{54 \phi_{k}^{2}}\right)^{1/3} - \frac{\phi_{k}^{2}}{4}\right] \left( \frac{\phi_{e}^{2}}{\phi_{k}^{2}}\right)\,.
	\end{equation}
	At the end of inflation, the scalar field reaches a value determined by the condition $ \epsilon_{1}= 1$. This value can be found using Eq. \eqref{eq15} as follows 
	\begin{equation}\label{eq35}
		\phi_{e} = \frac{1}{\left(\xi\left(1+3 \xi\right)\right)^{1/3}}
	\end{equation}
	The e-folding number during inflation, as expressed in Eq. \eqref{eq111}, is obtained as follows
	\begin{equation}\label{eq36}
		N_{k} = \frac{\xi \left(1+3 \xi \right)}{3} \left( \phi_{k}^{2}-\phi_{e}^{2}\right)\,,
	\end{equation}
	where $ \phi_{k} $ represents the inflaton value at the moment when the mode $ k $ exits the horizon, and $ \phi_{e} $ corresponds to the inflaton value at the end of inflation.
	By applying Eqs. \eqref{eq15} and \eqref{eq17}, $n_{s}$ can be determined as
	\begin{equation}
		n_{s}\simeq \frac{- 4.2+ 42.6 \phi^{3} \xi^{2}+\left( 14.2 \phi^{3}- 3.6\right) \xi}{\xi  \left( 1+ 3 \xi \right) \phi^{3}}\,.
	\end{equation}
	Next, by using the above equation, we find
	\begin{equation}\label{eq38}
		\phi_{k} = \left[ \frac{-3.6}{\xi \left( 3 n_{s} - 42.6\right)}\right]^{1/3}\,.
	\end{equation}
	Finally, by inserting Eqs. \eqref{eq35} and \eqref{eq38} into Eq. \eqref{eq36}, and assuming the condition $ \phi_{e} \ll \phi_{k} $, we obtain
	\begin{equation}
		N_{k} = \frac{-6 \xi}{5 n_{s} - 71}\,.
	\end{equation}
	To proceed further, we computed these values numerically and the reheating phase is analyzed by expressing $N_{re}$ (Eq. \eqref{eq30}) and $T_{re}$ (Eq. \eqref{eq31}) in terms of $n_{s}$, $A_{s}$, $\xi$, for various choices of $\omega_{re}$. In the left panel of Figure \eqref{fig.2}, we illustrate the behavior of $N_{re}$ as functions of $n_{s}$. Also, the right panel shows the behavior of $T_{re}$ as functions of $n_{s}$. We analyze a wide range for four cases of the specific values of $\omega_{re}$: $-1, -\frac{1}{3}, 0$ and $\frac{1}{6}$ that are consistent with the Planck2018 data. The cyan vertical band represents bounds on the scalar spectral index ($ n_{s}= 0.9658 \pm 0.0038 $) derived from recent astrophysical data such as Planck2018 data. Furthermore, we show in Fig. \eqref{fig.3} the allowed values for the number of e-folds during reheating $N_{re}$ and the equation of state parameter $\omega_{re} $. In the table \eqref{table-1}, we summarize the constraints of our model so that they are extracted from observationally consistent values of $n_{s}$ and \( r \) reported by the Planck2018 data.
	The common intersection point of all $\omega_{re} $ curves corresponds to the scenario of instantaneous reheating, where $N_{re} \rightarrow 0 $. At the end of the reheating phase, this temperature marks the highest value reached by the thermal bath.
	
	\begin{table}[H]
		\centering
		\caption{Constraints on the reheating phase parameters: the equation-of-state \( \omega_{re} \), the number of e-folds \( N_{re} \), and the reheating temperature \( T_{re} \) (\text{GeV}). These constraints shed light on the interplay between post-inflationary physics and final temperature.
		}
		\label{table-1}
		\setlength{\tabcolsep}{20pt}
		\begin{tabular}{ |c| c| c|}
			\hline
			\hline
			\rowcolor{Gray}	
			$ \omega_{re} $ & $N_{re}$ & $ T_{re} (\text{GeV}) $ \\
			\toprule
			$\frac{1}{6}$ & $13.32$ & $1.13\times10^{12}$ \\
			\hline
			$0$ & $6.64$ & $8.61\times10^{13}$\\
			\hline
			$-\frac{1}{3}$ & $3.32$ & $2.38\times10^{15}$\\
			\hline
			$-1$ & $1.65$ & $1.25\times10^{16}$\\
			\bottomrule	
		\end{tabular}
	\end{table}

	In our analysis, we chose the constant values of the effective equation-of-state parameter $\omega_{re}$ during reheating for simplicity and following the standard treatment in the literature. We note, however, that in realistic scenarios $\omega_{re}$ can vary with time during reheating, as pointed out in \cite{Saha:2020bis, Ghosh:2024ybs}. A more detailed treatment with time-dependent $\omega_{re}(t)$ is left for our future work.
	
	In the last step, we do an important check on the unitarity issue. In the standard Higgs inflation with a large non-minimal coupling ($\xi \sim 10^{4}$), the cutoff scale of the effective field theory is given by
		\begin{equation}
			\Lambda \sim \frac{M_{P}}{\xi}\,,
		\end{equation}
		which lies below the inflationary field values and may lead to unitarity issues \cite{Barbon:2009ya}. In that case, since the cutoff scale $\Lambda$ is lower than the field values relevant for inflation, it raises potential issues with perturbative unitarity and naturalness. In our setup, with a smaller coupling as $\xi = 6\times 10^{2}$, the cutoff is as fallows 
		\begin{equation}
			\Lambda \sim \frac{M_{P}}{\xi} \simeq \frac{2.4 \times 10^{18}\, \mathrm{GeV}}{6\times 10^{2}} \approx 4 \times 10^{15}\, \mathrm{GeV},
		\end{equation}
		 while by defining Eq. \eqref{eq32}, the Hubble scale during inflation in our model is obtained from
		 \begin{equation}
		 	H_{k} \;=\; \pi M_P \sqrt{\frac{r A_s}{2}}\,,
	 \end{equation}
 which, for typical parameter values and using the observational bound \(r=0.0017\), yields
 \begin{equation}
 	H_{k} \sim 1.022 \times10^{13} \, \mathrm{GeV}.
 \end{equation}
 Therefore, inflation occurs well below the cutoff ($ H_{k} < \Lambda $), ensuring that the effective theory remains valid throughout and the unitarity problem does not occur here. \\

	\section{Results}\label{sec-4}
	In this study, we investigated the Higgs inflation, and we employed the Higgs boson as the inflaton in the UG framework. We have shown that there exists the possibility of adding a non-minimal coupling that can cause inflation. Moreover, since UG is really a constrained model of GR, our findings remain consistent with conventional GR as well.
	In this setup, the NMC term helps smoothly move out of the inflationary phase without needing extra tools in unimodular non-minimal Higgs inflation.
	The main result of this work is a focus on the post-inflationary reheating era and a presentation of a comprehensive analysis of reheating within the framework of Higgs inflation. Our result emphasizes the importance of observational data in constraining theoretical models such as the UG Higgs cosmic inflation.
	We proceeded by extracting three key parameters that define the reheating phase: \( T_{re} \), \( \omega_{re} \), and \( N_{re} \). Then, we presented the results for four unique selections of the reheating equation of state parameter, namely, $\omega_{re} = -1, -\frac{1}{3}, 0, \frac{1}{6}$. Next, we computed the number of e-folds in the radiation dominated regime. In the last step, after considering all these analytical results, we obtained the reheating temperature \( T_{re} \). The analysis demonstrates that the model supports a viable reheating phase, from highly accelerated expansion phase to nearly radiation-dominated expansion. This allows us to probe how different physical assumptions impact on the thermal history of the universe. Finally, our results suggest that implementing Higgs inflation within the UG framework facilitates a viable reheating phase in this model. The results derived from the reheating epoch show good agreement with recent observational data, as confirmed by the Planck2018, BICEP/Keck2021, DESI2024, and ACT datasets.\\
	
	\appendix
\section{Supplementary Calculations}\label{appendix}
	Here we show equations and detailed calculations for slow roll parameters, perturbation parameters, and the number of e-folds. We work under the condition $\phi^{2} \gg \upsilon^{2}$, which allows us to approximate 
		\begin{equation}
		V=\frac{\gamma}{4}\left(\phi^{2}-\upsilon^{2}\right)^{2} \simeq \frac{\gamma}{4}\phi^{4}. 
	\end{equation} 
		We also define $\lambda = \phi^{2n}/(2n)$, with $n$ taken as an arbitrary power. Then by choosing $ n=2 $ and the slow-roll conditions, the field equations rewrite as 
	 \begin{equation}
			\begin{split}
				&\dot{\phi}\simeq -\frac{\left( 1+\gamma\right)a^{-6}}{12 \mathcal{H} \xi^{2}\left( 1+3 \xi \right) }\,,\\
				&\dot{\mathcal{H}}\simeq -\frac{\left( 1+\gamma\right) a^{-6}}{12  \xi^{2} \phi  \left(1+3 \xi \right)}-\frac{\left(1+\gamma \right) \phi^{2}a^{-6}}{4 \xi}\,,\\
				&\ddot{\phi}=\frac{\left( 1+\gamma\right) a^{-6}}{2\xi^{2} \left(1+3 \xi \right)}- \frac{a^{-6}}{12 \xi^{3} \phi^{3} \left(1+3 \xi \right)}-\frac{a^{-6}}{4 \xi^{2} \left(1+3 \xi \right)}\,,
			\end{split}
		\end{equation}
		Here we set $ M_{P}=1$, approximate $1+\gamma \simeq 1$, and use $\phi^{2}\gg M^{2}_{P}/\xi$ as dominant.
		In this regime, we calculate slow-roll parameters as follows
		\begin{equation}
			\begin{split}
				&\epsilon_{1} \simeq\frac{1}{\xi \phi^{3} \left(1+3 \xi \right)}\,, \quad \epsilon_{2} \simeq \frac{1+\left(-9 \gamma -6\right) \xi  \,\phi^{3}}{\xi \phi^{3}\left(1+\gamma \right)}\,, \\
				&\epsilon_{3} \simeq -\frac{1}{\xi\phi^{3} \left(1+3 \xi \right)}\,, \quad \epsilon_{4} \simeq	-\frac{1}{\xi \phi^{3} \left(1+3 \xi \right)}\,,
			\end{split}
		\end{equation}
		At this point, by using \eqref{eq111}, the number of e-folds obtained is $ N=\xi\left(1+3 \xi \right)\frac{\phi^{3}}{3} \bigg|_{\phi_{e}}^{\phi_{k}} $. Therefor, by approximate $ \phi_{e} \ll \phi_{k} $, we have $ \phi_{k}\simeq \left(\frac{3N}{\xi\left(1+3\xi \right)} \right)^{1/3} $. Now the perturbation parameters are defined as fallows
		\begin{equation}
			n_{s}\simeq \frac{\left(- 4.2+\frac{ 127.8 N \xi}{1+3 \xi}+\left(\frac{ 42.6 N}{\xi  \left(1+3 \xi \right)}- 3.6\right) \xi \right) \left(1+3 \xi \right)}{3 \left( 1.0+ 3.0 \xi \right) N}\,,
		\end{equation}
		and
		\begin{equation}
			r\simeq \frac{32+192 \xi}{\xi  \left(\frac{18 N \xi}{1+3 \xi}+\frac{6 N}{1+3 \xi}-1\right)^{2}}\,.
		\end{equation}
		Our analysis with \textit{Maple} yielded three viable scenarios consistent with the observational data for $n_{s}$ and $r$ (i.e., $\xi= 6 \times 10^{2}, 6.01 \times 10^{2}, 6.1 \times 10^{2}$). Then we calculated for these three parameters for various values of $ N $. We find out that for $\xi= 6 \times 10^{2}$ and $ N= 55 $ there is good agreement with observational data.\\

\end{document}